\documentclass[twocolumn,showpacs,preprintnumbers,amsmath,amssymb]{revtex4}
\usepackage{graphicx}% Include figure files
\usepackage{dcolumn}% Align table columns on decimal point
\usepackage{bm}% bold math
\usepackage{verbatim}
\usepackage{color}
\usepackage[colorlinks]{hyperref}
\input{epsf}
\begin{document}

%\preprint{APS/123-QED}

\title{Microwave-induced electron heating in the regime of radiation-induced magnetoresistance oscillations}

\author{A. N. Ramanayaka}
% \altaffiliation{Department of Physics and Astronomy, Georgia State University,
%Atlanta, GA 30303.}%Lines break automatically or can be forced with \\
\author{R. G. Mani}
\affiliation{Georgia State University, Atlanta, GA 30303.}

\author{W. Wegscheider}
\affiliation{Laboratorium f\"{u}r Festk\"{o}rperphysik, ETH
Z\"{u}rich, 8093 Z\"{u}rich, Switzerland}

\date{\today}% It is always \today, today,

\begin{abstract}
We examine the influence of microwave photoexcitation on the
amplitude of Shubnikov-de Haas (SdH) oscillations in a two
dimensional GaAs/AlGaAs electron system in a regime where the
cyclotron frequency, $\omega_{c}$, and the microwave angular
frequency, $\omega$, satisfy $2 \omega \le \omega_{c} \le 3.5
\omega$. A SdH lineshape analysis indicates that increasing the
incident microwave power has a weak effect on the amplitude of the
SdH oscillations, in comparison to the influence of modest
temperature changes on the dark-specimen SdH effect. The results
indicate negligible electron heating under modest microwave
photoexcitation, in good agreement with theoretical predictions.
\end{abstract}

\maketitle

\section{introduction}
The GaAs/AlGaAs quasi two-dimensional electron system has served
as the basis for many interesting developments in modern condensed
matter physics.\cite{grid-2,grid-1} In the recent past,
photo-excited transport studies in this system have become a topic
of experimental\cite{grid1, grid2, grid101, grid3, grid4, grid5,
grid6, grid7, grid8, grid9, grid10, grid11, grid12, grid13,
grid14, grid15, grid16, grid17, grid18, grid19, grid20, grid21,
grid22} and theoretical\cite{grid23, grid24, grid25, grid26,
grid27, grid28, grid29, grid30, grid31, grid32, grid33, grid34,
grid35, grid36, grid37, grid38, grid39, grid40, grid41, grid42,
grid43, grid44} interest since the observation of zero-resistance
states and associated magneto-resistance oscillations in the
microwave excited two-dimensional electron system.\cite{grid1,
grid2}. Periodic in $B^{-1}$ radiation-induced magnetoresistance
oscillations, which lead into the radiation-induced
zero-resistance states, are now understood to be a consequence of
radiation-frequency ($f$), and magnetic field ($B$) dependent,
scattering at impurities \cite{grid23, grid25, grid26} and/or a
change in the distribution function,\cite{grid32, grid101} while
vanishing resistance is thought to be an outcome of negative
resistance instability and current domain
formation.\cite{grid24,grid41} Although there has been much
progress in this field, there remain many aspects, such as
indications of activated transport, the overlap with quantum Hall
effect, and the influence of the scattering lifetimes, that could
be better understood from both the experimental and theoretical
perspectives.

A further topic of experimental interest is to examine the
possibility of electron heating, as theory has,\cite{grid28,
grid31, grid33} in consistency with common experience, indicated
the possibility of microwave-induced electron heating in the high
mobility 2DES in the regime of the radiation-induced
magnetoresistance oscillations. Not surprisingly, under steady
state microwave excitation, the 2DES can be expected to absorb
energy from the radiation field. At the same time, electron-phonon
scattering can serve to dissipate this surplus energy onto the
host lattice. Lei et. al \cite{grid31} have determined the
electron temperature, $T_{e}$, by balancing the energy dissipation
to the lattice and the energy absorption from the radiation field,
while including both intra-Landau level and inter-Landau level
processes. In particular, they showed that the electron
temperature, $T_{e}$, the longitudinal magnetoresistance,
$R_{xx}$, and the energy absorption rate, $S_{p}$, can exhibit
remarkable correlated non-monotonic variation vs.
$\omega_{c}/\omega$, where $\omega_{c}$ is the cyclotron
frequency, and $\omega = 2 \pi f$, with $f$ the radiation
frequency.\cite{grid31} In such a situation, some questions of
experimental interest then are: (a) How to probe and measure
electron heating in the microwave-excited 2DES? (b) What is the
magnitude of electron heating under typical experimental
conditions? Finally, (c) is significant electron heating a general
characteristic in microwave radiation-induced transport?

An approach to the characterization of electron-heating could
involve a study of the amplitude of the Shubnikov-de Haas (SdH)
oscillations, that also occur in $R_{xx}$ in the photo-excited
specimen. Typically, SdH oscillations are manifested at higher
magnetic fields, $B$, than the radiation-induced magnetoresistance
oscillations, i.e., $B > B_{f} = 2\pi f m^{*}/e$, especially at
low microwave frequencies, say $f \le 50 GHz$ at $T \ge 1.3K$. On
the other hand, at higher $f$, SdH oscillations can extend into
the radiation-induced magneto-resistance oscillations. In a
previous study, ref.\cite{grid16} has reported that SdH
oscillation amplitude scales linearly with the average background
resistance in the vicinity of the radiation-induced resistance
minima, indicating the SdH oscillations vanish in proportion to
the background resistance at the centers of the radiation-induced
zero-resistance states. Kovalev et. al \cite{grid7} have reported
the observation of a node in the SdH oscillations at relatively
high-$f$. Ref. \cite{grid14} discuss SdH damping and a strong
suppression of magnetoresistance in a regime where microwaves
induce intra-Landau-level transitions. Both ref.\cite{grid16} and
ref. \cite{grid7} examined the range of $\omega_{c}/\omega \leq
1$, whereas ref.\cite{grid14} examined the $\omega_{c}/\omega \geq
1$ regime.

From the theoretical perspective, Lei et al. have suggested that a
modulation of SdH oscillation amplitude in $R_{xx}$ results from
microwave-electron heating. Further, they have shown that, in
$\omega_{c}/\omega \leq 1$ regime, both $T_{e}$ and $S_{p}$
exhibit similar oscillatory features, while in $\omega_{c}/\omega
\geq 1$ regime, both $T_{e}$ and $S_{p}$ exhibit a relatively flat
response.

%\begin{comment}
\begin{figure}[t]
%h=here, t=top, b=bottom, p=separate figure page
\begin{center}
\leavevmode \epsfxsize=2.5 in \epsfbox {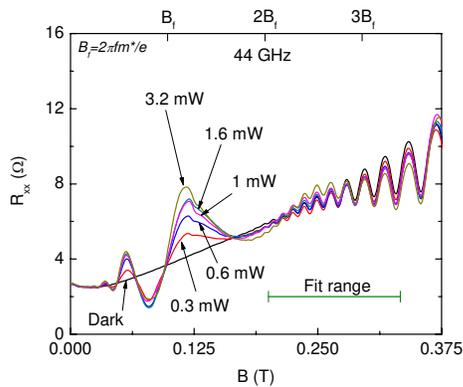}
\end{center}
%\begin{figure}
%\includegraphics{fig1}% Here is how to import EPS art
\caption{\label{fig:epsart} (color online). Microwave induced
magneto-resistance oscillations  and SdH oscillations in $R_{xx}$
are shown at $1.5 K$ for $44 GHz$ at different power levels, $P$.
A horizontal marker (green) shows the field range ($2B_{f} \le B
\le 7/2 B_{f}$) where SdH fits were carried out.}

\end{figure}

\begin{figure}[t]
%h=here, t=top, b=bottom, p=separate figure page
\begin{center}
\leavevmode \epsfxsize=2.5 in \epsfbox {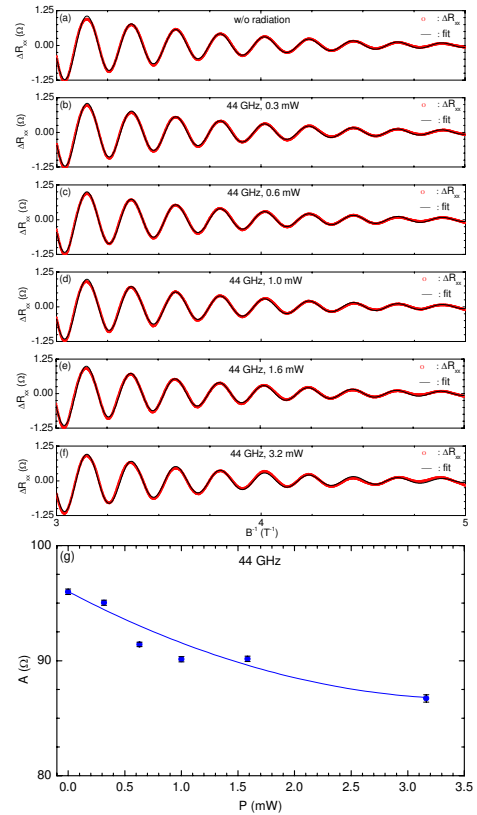}
\end{center}
%\begin{figure}
%\includegraphics{fig1}% Here is how to import EPS art
\caption{\label{fig:epsart} (color online). (a) The background
subtracted $R_{xx}$, i.e., $\Delta R_{xx}$, in the absence of
radiation (open circles) and a numerical fit (solid line) to
$\Delta R_{xx}$ = -$ A e^{-\alpha /B} \cos(2 \pi F/B)$ are shown
here. Panels (b) $-$ (f) show the $\Delta R_{xx}$ and the fit at
the indicated $P$. Panel (g) shows $A$ vs. $P$ at $44 GHz$.}
\end{figure}

\begin{figure}[t]
%h=here, t=top, b=bottom, p=separate figure page
\begin{center}
\leavevmode \epsfxsize=2.5 in \epsfbox {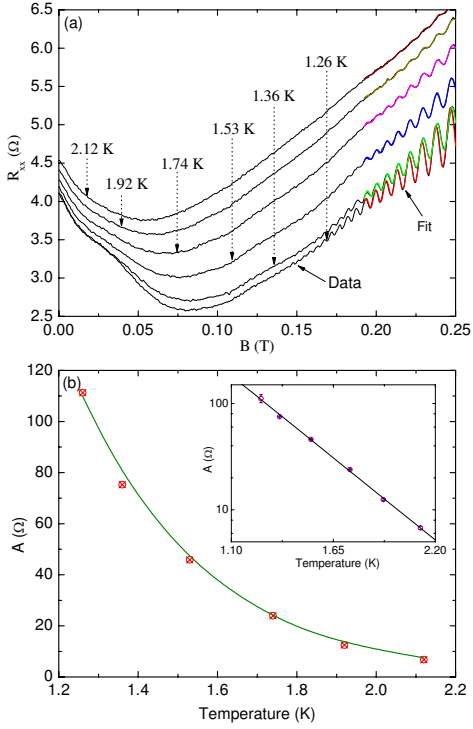}
\end{center}
%\begin{figure}
%\includegraphics{fig1}% Here is how to import EPS art
\caption{\label{fig:epsart} (color online).(a) Temperature ($T$)
dependence of $R_{xx}$ is shown for $1.26K \le T \le 2.12K$. The
black lines show the data, and the colored lines indicate the
fits. (b) The exponential variation of the amplitude $A^{\prime}$
with T is shown here. The inset confirms that $log(A^{\prime})$ is
linear in T.}
\end{figure}

\begin{figure}[t]
%h=here, t=top, b=bottom, p=separate figure page
\begin{center}
\leavevmode \epsfxsize=2.5 in \epsfbox {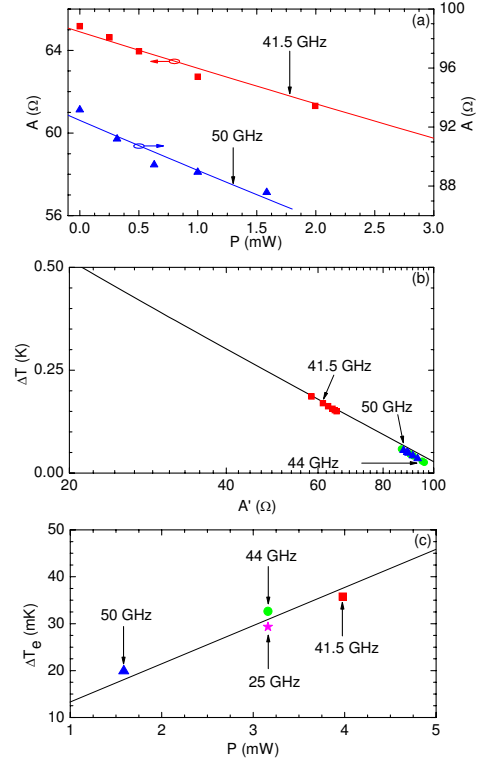}
\end{center}
%\begin{figure}
%\includegraphics{fig1}% Here is how to import EPS art
\caption{\label{fig:epsart} (color online). (a) Variation of $A$
with  $P$ is shown for $f = 41.5 GHz$ (left axis) and $f = 50 GHz$
(right axis).  (b) The effect of $T$ and $P$ on $A^{\prime}$
(solid line) and $A$ (solid symbols) are shown here. Corresponding
$T$-change was determined through the exponential relation between
$\Delta T$ and $A^{\prime}$. Panel (c) shows a plot of maximum
$\Delta T_{e}$ vs peak-$P$ (solid symbols), with a line to guide
the eye.}

\end{figure}

%\end{comment}

Here, we investigate the effect of microwaves on the SdH
oscillations over $2B_{f} \le B \le 3.5 B_{f}$, i.e., $2 \omega
\le \omega_{c} \le 3.5 \omega$, where $B_{f}=2 \pi f m^{*}/e$,
$m^{*}$ is the effective electron mass, and $e$ is the electron
charge\cite{grid1}. In particular, we compare the relative change
in the SdH oscillation amplitude due to lattice temperature
changes in the dark, with changes in the SdH amplitude under
microwave excitation at different microwave power levels, at a
constant bath temperature. From such a study, we extract the
change in the electron temperature, $\Delta T_{e}$, induced by
microwaves. In good agreement with theory, the results indicate
$\Delta T_{e} \le 50 mK$ over the examined regime.

\section{Experiment and Results}

The lock-in based electrical measurements were performed on Hall
bar devices fabricated from high quality GaAs/AlGaAs
heterostructures. Experiments were carried out with the specimen
mounted inside a waveguide and immersed in pumped liquid helium.
The frequency spanned $25 \le f \le 50 GHz$ at source power levels
$P \leq 5 mW$. Magnetic-field-sweeps of $R_{xx}$ vs. $P$ were
carried out at $1.6 K$ at $41.5GHz$, and at $1.5 K$ at $44 GHz$
and $50 GHz$.

Microwave-induced magneto-resistance oscillations can be seen in
Fig. 1 at $B \le 0.175$ T, as strong SdH oscillations are also
observable under both the dark and irradiated conditions for $B
\ge 0.2 T$. Over the interval $2B_{f} \le B \le 3.5 B_{f}$, where
the SdH oscillations are observable, one observes small variations
in the background $R_{xx}$ at higher power levels. Thus, a smooth
$R_{xx}$ background was subtracted from the magneto-resistance
data. Figure 2 (a) - (f) shows the background subtracted $R_{xx}$,
i.e., $\Delta R_{xx}$, measured without (Fig. 2(a)) and with (Fig.
2(b)-2(f)) microwave radiation versus the inverse magnetic field,
$B^{-1}$. To extract the amplitude of the SdH oscillations, we
performed a standard Nonlinear Least Square Fit (NLSF) on $\Delta
R_{xx}$ data with a exponentially damped sinusoid, i.e., $\Delta
R_{xx} = -A e^{-\alpha /B} \cos(2 \pi F /B)$. Here, $A$ is the
amplitude, $F$ is the SdH frequency, and $\alpha$ is the damping
factor. The fit results for the dark-specimen $\Delta R_{xx}$ data
are shown in the Fig. 2 (a) as a solid line. This panel suggests
good agreement between data and fit in the dark condition.
Similarly, we performed NLSFs of the $\Delta R_{xx}$ SdH data
taken with the microwave power spanning approximately $0 \le P \le
3mW$, see Fig. 2(b)- 2(f). Since the parameters $\alpha$ and $F$
are insensitive to the incident radiation at a constant
temperature, we fixed $\alpha$ and $F$ to the dark-specimen
constant values. In Fig. 2, panels (b) $-$ (f) show the $T = 1.5K$
$\Delta R_{xx}$ data (open circles) and fit (solid line) obtained
with $f = 44GHz$ for different power levels. The SdH amplitude $A$
extracted from the NLSFs are exhibited vs. the microwave power in
Fig. 2(g). Here, $A$ decreases with increasing microwave power.
Our analysis of other power-dependent data (not shown) yielded
similar results.

Next, we examine the influence of temperature on the SdH
oscillation amplitude. Thus, Fig.3(a) shows $R_{xx}$ vs. $B$ with
the temperature as a parameter. It is clear, see the black solid
lines, that increasing the temperature rapidly damps the SdH
oscillations at these low magnetic fields. In order to extract the
SdH amplitude from these data, we used the same fitting model,
$\Delta R_{xx} = -A e^{-\alpha /B} \cos(2 \pi F/B)$, as described
previously. But, for the $T$-dependence analysis, $\alpha$ was
separated into two parts, $\alpha_{T_{0}}$ and $\beta \Delta T$,
i.e., $\alpha = \alpha_{T_{0}} + \beta \Delta T$, since we wished
to relate the change in the SdH amplitude for a temperature
increment to the observed change in the SdH amplitude for an
increment in the microwave power at a fixed $f$. Here
$\alpha_{T_{0}}$ represents the damping at the base temperature,
and $\beta \Delta T$ is the additional damping due to the
temperature increment, $\Delta T$ = $T - T_{0}$. Now the fit
function becomes $\Delta R_{xx}$ = $ -A e^{- ( \alpha_{T_{0}} +
\beta \Delta T) /B} \cos(2 \pi F/B)$ = -$A^{\prime} e^{-
\alpha_{T_{0}/B}} \cos(2 \pi F/B)$. Here, $ A^{\prime} = A e^{-
\beta \Delta T /B}$ represents the change in the SdH oscillation
amplitude due to the $\Delta T$. Note that the parameters
$\alpha_{T_{0}}$ and $F$ can be extracted from the $R_{xx}$ fit at
the lowest $T$ and set to constant values. Data fits to $\Delta
R_{xx}$ = -$A^{\prime} e^{- \alpha_{T_{0}/B}} \cos(2 \pi F/B)$ are
included in Fig. 3(a) as colored solid lines. Thus, the NLSF
served to determine $A^{\prime}$ at each temperature. Fig. 3(b)
shows the temperature dependence of $A^{\prime}$, while the inset
of Fig. 3(b) shows a semi-log plot of $A^{\prime}$ vs. $T$. The
inset confirms an exponential dependence for $A^{\prime}$ on $T$,
i.e, $ A^{\prime} = A e^{- \beta \Delta T /B}$.

Experiments indicate that increasing the source-power
monotonically decreases the SdH oscillation amplitude at the
examined frequencies including $44 GHz$ (see Fig.2(g)), $50 GHz$
and $41.5 GHz$ (see Fig.4(a)). Since increasing the temperature
also decreases the SdH oscillation amplitude (see Fig.3(b)), one
might extract the electron temperature change under microwave
excitation by inverting the observed relationship between
$A^{\prime}$ and $T$, i.e., since $ A^{\prime} = A e^{- \beta
\Delta T /B}$, $\Delta T = - (B/\beta^{-1})(lnA^{\prime} + c)$,
where $\beta$ is a constant. Thus, the dark measurement of the SdH
amplitude vs. the temperature serves to calibrate the temperature
scale vs. the SdH amplitude, and the slope of the solid line in
Fig. 4(b) reflects the inverse slope of Fig. 3(b). Also plotted as
solid symbols in Fig.4(b) are the $A$ under microwave excitation
at various frequencies and power levels. Here, one observes that
the change in SdH amplitude induced by microwave excitation over
the available power range is significantly smaller than the change
in SdH amplitude induced by a temperature change of 0.9K. By
transforming the observed change in $A$ between minimum- and
maximum- power at each $f$ to a $\Delta T_{e}$, we can extract the
maximum $\Delta T_{e}$ induced by photo-excitation at each $f$,
and this is plotted in Fig.4(c). Although  the power at the sample
can vary with $f$, even at the same source power,
Fig.4(c)indicates that the maximum $\Delta T_{e}$ scales
approximately linearly with the peak source microwave power (see
Fig.4(c) solid guide line). From Fig.4(c), it appears that $\Delta
T_{e}/\Delta P = 9 mK/mW$.

\section{Discussion}

According to theory,\cite{grid28,grid31,grid33}steady state
microwave excitation can heat a high mobility two-dimensional
electron system. The energy gain from the radiation field is
balanced by energy loss to the lattice by electron-phonon
scattering. In ref. \cite{grid31}, Lei et al. suggest that
longitudinal acoustic (LA) phonons provide for more energy
dissipation than the transverse acoustic (TA) phonons in the
vicinity of $T = 1K$ in the GaAs/AlGaAs system, if one neglects
the surface or interface phonons. At such low temperatures and
modest microwave power, away from the cyclotron resonance
condition, where the resonantly absorbed power from the microwave
radiation is not too large, and the electron temperature remains
well below about $20K$, the longitudinal optical(LO) phonons do
not influence the resistivity since the energy scale associated
with LO-phonons is large compared to the energy scale for acoustic
phonons. Within their theory, Lei et al.\cite{grid31} show that
the electron temperature follows the absorption rate, exhibiting
rapid oscillatory behavior at low $B$, i.e., $\omega_{c}/ \omega
\leq 1$, followed by slower variation at higher $B$, i.e.,
$\omega_{c}/ \omega \geq 1$. Indeed, even at the very high
microwave intensities, $P/A$, e.g. $4 \le P/A \le 18 mW/cm^{2}$ at
$50 GHz$, theory does not show a significant change in the
electron temperature for $B \ge 2B_{f}$ except at $P/A \geq 10.5
mW/cm^{2}$. Thus, theory indicates that the electron temperature
is nearly the lattice temperature in low $P/A$ limit especially
for $B \ge 2B_{f}$, i.e., $\omega_{c} \ge 2 \omega$. In
comparison, in these experiments, the source power satisfied $P
\le 4 mW$, with $A \approx 1 cm^{2}$, while the power at the
sample could be as much as ten times lower due to attenuation by
the hardware. Remarkably, we observe strong microwave induced
resistance oscillations (see Fig.1) in the $B \le B_{f}$ regime at
these low intensities. The results, see Fig. 4(c), indicate just a
small rise, $\Delta T \le 50 \times 10^{-3}K$, in the electron
temperature, $T_{e}$, above the lattice temperature, $T$.

\section{Summary and Conclusions}

In summary, this experimental study indicates that the perceptible
effect of the incident microwave radiation on the amplitude of the
SdH oscillations over the regime $2 \omega \le \omega_{c} \le 3.5
\omega$, when the microwave excitation is sufficient to induce
strong microwave induced resistance oscillations, corresponds to a
relatively small increase in the electron temperature, in good
agreement with theoretical predictions.

This work has been supported by D. Woolard and the ARO under
W911NF-07-01-0158, and by A. Schwartz and the DOE under
DE-SC0001762.


\begin{thebibliography}{53}
\bibitem{grid-2} R. E. Prange and S. M. Girvin, The Quantum Hall
Effect, 2nd. ed. (Springer, New York, 1990).

\bibitem{grid-1} S. Das Sarma and A. Pinczuk, Perspectives in
Quantum Hall Effects (Wiley, New York, 1996).

\bibitem{grid1} R. G. Mani, J. H. Smet, K. von Klitzing, V. Narayanamurti, W.
B. Johnson, and V. Umansky, Nature (London)  420, 646 (2002).

\bibitem{grid2} M. A. Zudov, R. R. Du, L. N. Pfeiffer, and K. W. West, Phys.
Rev. Lett. 90, 046807 (2003).

\bibitem{grid101} S. Dorozhkin, JETP Lett. \textbf{77}, 577 (2003).

\bibitem{grid3}R. G. Mani, V. Narayanamurti, K. von Klitzing, J. H. Smet, W. B. Johnson, and V. Umansky, Phys. Rev. B 69, 161306 (2004); 70, 155310
(2004).

\bibitem{grid4} R. G. Mani et al., Phys. Rev. Lett. 92, 146801 (2004);
Phys. Rev. B 69, 193304 (2004).

\bibitem{grid5} R. G. Mani, Physica E (Amsterdam) 22, 1 (2004); ibid. 25, 189
(2004).

\bibitem{grid6} R. G. Mani, Appl. Phys. Lett. 85, 4962 (2004).

\bibitem{grid7} A. E. Kovalev et al., Solid State Commun. 130, 379 (2004).

\bibitem{grid8} S. A. Studenikin et al., Solid State Commun. 129, 341 (2004).

\bibitem{grid9} R. R. Du et al., Physica E (Amsterdam) 22, 7
(2004).

\bibitem{grid10} I. V. Kukushkin et al., Phys. Rev. Lett. \textbf{92}, 236803
(2004).

\bibitem{grid11} R. G. Mani, IEEE Trans. Nanotechnol. 4, 27 (2005); Phys. Rev. B 72, 075327 (2005); Sol. St. Comm. 144, 409 (2007); Appl. Phys. Lett. 92, 102107 (2008); Physica E 40, 1178
(2008).

\bibitem{grid12} B. Simovic et al., Phys. Rev. B 71, 233303
(2005).

\bibitem{grid13} J. H. Smet et al., Phys. Rev. Lett. 95, 116804
(2005).

\bibitem{grid14} S. I. Dorozhkin et al., Phys. Rev. B 71, 201306(R)
(2005).

\bibitem{grid15} Z. Q. Yuan et al., Phys. Rev. B 74, 075313 (2006).

\bibitem{grid16} R. G. Mani, Appl. Phys. Lett. 91, 132103 (2007).

\bibitem{grid17} S. A. Studenikin et al., Phys. Rev. B 76, 165321
(2007).

\bibitem{grid18} K. Stone et al., Phys. Rev. B 76, 153306 (2007).

\bibitem{grid19} A. Wirthmann et al., Phys. Rev. B 76, 195315 (2007).

\bibitem{grid20} S. Wiedmann et al., Phys. Rev. B 78, 121301(R)
(2008).

\bibitem{grid21} R. G. Mani et al., Phys. Rev. B 79, 205320
(2009); ibid. 81, 125320 (2010).

\bibitem{grid22} O. M. Fedorych et al., Phys. Rev. B 81, 201302 (2010).

\bibitem{grid23} A. C. Durst et al., Phys. Rev. Lett. 91, 086803 (2003).

\bibitem{grid24} A. V. Andreev et al., Phys. Rev. Lett.
91, 056803 (2003).

\bibitem{grid25} V. Ryzhii and A. Satou, J. Phys. Soc. Jpn. 72, 2718 (2003).

\bibitem{grid26} X. L. Lei and S. Y. Liu, Phys. Rev. Lett. 91, 226805 (2003).

\bibitem{grid27} P. H. Rivera, P. A. Schulz, Phys. Rev. B 70, 075314
(2004).

\bibitem{grid28} X. L. Lei, J. Phys.: Condens. Matter 16, 4045 (2004).

\bibitem{grid29} S. A. Mikhailov, Phys. Rev. B 70, 165311 (2004).

\bibitem{grid30} J. Inarrea and G. Platero, Phys. Rev. B 72, 193414 (2005).

\bibitem{grid31} X. L. Lei and S. Y. Liu, Phys. Rev. B 72, 075345 (2005).

\bibitem{grid32} I. A. Dmitriev et al., Phys. Rev. B 71, 115316 (2005).

\bibitem{grid33} J. Inarrea and G. Platero, Phys. Rev. Lett. 94, 016806 (2005).

\bibitem{grid34} A. Auerbach et al., Phys. Rev.
Lett. 94, 196801 (2005).

\bibitem{grid35} J. Inarrea and G. Platero, Appl. Phys. Lett. 89, 052109 (2006); ibid. 89, 172 114 (2006);
ibid. 90, 172118 (2007); ibid. 90, 262101 (2007); ibid. 92, 192113
(2008); Phys. Rev. B 80, 193302 (2009).

\bibitem{grid36} J. Inarrea and G. Platero, Phys. Rev. B 76, 073311 (2007); ibid. 78, 193310 (2008).

\bibitem{grid37} A. D. Chepelianskii et al., Eur.
Phys. J. B 60, 225 (2007).

\bibitem{grid38} A. Auerbach and G. V. Pai, Phys. Rev. B 76, 205318 (2007).

\bibitem{grid39} I. A. Dmitriev et al., Phys. Rev. B 75,
245320 (2007).

\bibitem{grid40} P. H. Rivera et al., Phys. Rev. B 79, 205406 (2009).

\bibitem{grid41} I. G. Finkler and B. I. Halperin, Phys. Rev. B 79, 085315 (2009).

\bibitem{grid42} X. L. Lei and S. Y. Liu, Appl. Phys. Lett. 94,
232107 (2009).

\bibitem{grid43} A. D. Chepelianskii and D. L. Shepelyansky, Phys. Rev. B 80,
241308(R) (2009).


\bibitem{grid44}D. Hagenmuller et al.,
Phys. Rev. B 81, 235303 (2010).




\end{thebibliography}
\end{document}